\documentclass{aa}
\usepackage{txfonts}
\usepackage{graphicx}

\usepackage{color}

\begin{document}

\title{First deep search of tidal tails in the Milky Way globular cluster NGC~6362}

\author{Andr\'es E. Piatti\inst{1,2}\thanks{\email{andres.piatti@fcen.uncu.edu.ar}}}

\institute{Instituto Interdisciplinario de Ciencias B\'asicas (ICB), CONICET-UNCUYO, Padre J. Contreras 1300, M5502JMA, Mendoza, Argentina;
\and Consejo Nacional de Investigaciones Cient\'{\i}ficas y T\'ecnicas (CONICET), Godoy Cruz 2290, C1425FQB,  Buenos Aires, Argentina\\
}

\date{Received / Accepted}

\abstract{I present results of the analysis of a set of images obtained in the field of 
the Milky Way globular cluster NGC~6362 using the Dark Energy Camera, which is mounted in 
the 4.0m Victor Blanco telescope of the Cerro-Tololo Interamerican observatory. The cluster was
selected as a science case for deep high-quality photometry because of the controversial 
observational findings and theoretical predictions on the existence of cluster tidal tails. 
The collected data allowed us to build an unprecedented deep cluster field color-magnitude diagram,
from which I filtered stars to produce a stellar density map, to trace the stellar density 
variation as a function of the position angle for different concentric annulii centered on the
cluster, and to construct a cluster stellar density radial profile. I also built a stellar density map from
a synthetic color-magnitude diagram generated from a model of the stellar population distribution 
in the Milky Way. All the analysis approached converge toward a relatively smooth stellar density
between 1 and $\sim$ 3.8 cluster Jacobi radii, with a slightly difference smaller 
than 2 times the background stellar density fluctuation between the mean stellar density of the 
south-eastern and that of north-western hemispheres, the latter being higher. Moreover,
the spatial distribution of the recently claimed tidal tail stars agrees well not
only with the observed composite star field distribution, but also with the region
least affected by interstellar absorption. Nevertheless, I detected
 a low stellar density excess around the cluster Jacobi radius, from which
I conclude that NGC~6362 present a thin extra tidal halo.} 
 

\titlerunning{The tidal tails of NGC~6362}

\authorrunning{Andr\'es E. Piatti}

\maketitle

\markboth{Andr\'es E. Piatti: }{The tidal tails of NGC6362}

\section{Introduction}

The recent stringent compilation of Milky Way globular clusters with robust 
detections of extra-tidal structures includes only one bulge/disk globular cluster 
with tidal tails \citep[NGC~6362,][]{zhangetal2022}; while \citet{cg2021}
do not include any bulge globular cluster in their search for dark
matter in the globular clusters' outskirts. This brief overview illustrates that
bulge globular clusters have not generally been targeted for studies of their
outermost stellar structures, which are fundamental for our understanding
of whether they formed in dark matter minihaloes \citep{starkmanetal2020,bv2021,wanetal2021}; 
their association to destroyed dwarf progenitors \citep{carballobelloetal2014,mackeyetal2019};
their dynamical history as a consequence of the interaction with the Milky Way 
\citep{hb2015,deboeretal2019,pcb2020}, etc. Therefore, there are strong motivations 
for detecting/characterizing tidal tails in bulge globular clusters, making it a compelling 
field of research.

\citet{zhangetal2022} classified NGC~6362 as a globular cluster with tidal tails 
based on the work by \citet{kunduetal2019b}, who found 259 of them spread
over an area of $\sim$4.1 deg$^2$ centered on the cluster. These stars should be placed
beyond the cluster's Jacobi radius ($r_J$), which defines the 
surface from which cluster stars are not longer bounded to the cluster and are lost in the form 
of tidal tails. The value of  $r_J$ depends on the Milky Way potential at the position of the 
globular and its orbit around the Milky Way. For NGC~6362, \citet{piattietal2019b}
derived $r_J$ = 0.26 deg. \citet{kunduetal2019b} also showed that the
cluster is moving in a chaotic orbit. However, \citet{mestreetal2020} compared the behavior 
of simulated streams embedded 
in chaotic and non-chaotic regions of the phase-space and found that typical 
gravitational potentials of host galaxies can sustain chaotic orbits, which in turn 
do reduce the time interval during which streams can be detected. This explains
why tidal tails in some globular clusters are washed out afterwards they are generated 
to the point at which it is impossible to detect them. Indeed, 
NGC~5139, with an apogalactocentric distance of 7.0 kpc, is the innermost globular cluster 
with observed tidal tails; NGC~6362 is at 5.5 kpc from the Galactic center \citep{bv2021}. 

Given that the structures I am interested in --the tidal tails of NGC~6362-- are mainly 
composed of low-mass Main Sequence stars, it is necessary to map the faint end of the cluster 
color-magnitude diagram (CMD), allowing us to determine the outer structure 
of the cluster with excellent statistics and to truly map the outskirts to look for tidal 
tails. It is widely accepted that Milky Way globular clusters have lost most of their 
masses through three main processes, namely: stellar evolution, two-body relaxation and tidal 
heating caused by the Milky Way's gravitational field 
\citep[][and reference therein]{piattietal2019b}. Studies on the external regions of Milky Way 
globular clusters use such faint Main Sequence stars to detect tidal tails, because they are 
more numerous  \citep{carballobelloetal2012}. I note that the
{\it Gaia} data used by \citet{kunduetal2019b} to select tidal tail stars barely reach
the cluster's Main Sequence turnoff ($G_0$ $\sim$ 19 mag).

In Section 2 I describe the deep wide-field observations carried out to investigate the
existence of tidal tails in the outskirts of NGC~6362. The analysis of the obtained CMD and 
stellar density maps is described in Section 3, while Section 4 deals with the discussion of 
the present outcomes to the light of our knowledge of the distribution of stars and gas
in the Milky way along the cluster's line-of-sight. In Section 5 I summarize the main
conclusions of this work.

\section{Observational data}

I employed the Dark Energy Camera \citep[DECam,][]{flaugheretal2015}, an array of 62 identical
chips with a scale of 0.263 arcsec pixel$^{-1}$ that provides a 3 deg$^2$ field of view, attached
to the prime focus of the 4-m Blanco telescope at the Cerro Tololo Inter-American Observatory
(CTIO). The collected data are part of the observing program 2023A-627924 (PI: A. Piatti) and
consist of 4$\times$400s $g$ and 1$\times$100s + 7$\times$400s $i$ exposures, respectively. 
The images' quality resulted to be better than 0.9 arcsec.
The DECam community pipeline team processed the images by applying the highest performance instrumental
calibrations, eliminating one CCD of problematic usefulness. The CCD images were trimmed
of bad edge pixels, and reference bias and dome flat files were created for each night in order
to remove the instrumental signature from the science data. The images were then resampled 
to a standard orientation and pixel scale at a standard tangent point. Finally, the
DECam community pipeline applied photometric calibrations.

I split the entire DECam field of view in nine equal squared areas of $\sim$ 0.67 deg a side, and
for each of these sub-fields I performed point spread function (PSF) photometry using the 
{\sc daophot/allstar} suite of programs \citep{setal90}. For each sub-field, I obtained a quadratically 
varying PSF by fitting $\sim$ 2200 stars deg$^2$, once I eliminated the neighbors using
a preliminary PSF derived from the brightest, least contaminated
$900$ stars deg$^2$. Both groups of PSF stars were interactively selected. We
then used the {\sc allstar} program to apply the resulting PSF to the
identified stellar objects and to create a subtracted image, which was
used to find and measure magnitudes of additional fainter stars. This
procedure was repeated three times for each sub-field. In order to properly match
a $g$ sub-field with the corresponding $i$ one, I used the {\sc iraf.immatch@wcsmap} task,
and then stand-alone {\sc daomatch/daomaster} routines\footnote{programs kindly provided by P.B.
Stetson} to gather pixel coordinates, $g$ and $i$
magnitudes with their respective errors, sharpness and  $\chi$ values for all the measured 
sources.  Finally, I assigned R.A. and Dec. coordinates to each stars using the
{\sc wcstool} package\footnote{http://tdc-www.harvard.edu/wcstools/} and kept sources with
$|$sharpness$|$ $<$ 0.5 in order to remove bad pixels, unresolved double stars, cosmic
rays, and background galaxies from the photometric catalog.

\begin{figure}
\includegraphics[width=\columnwidth]{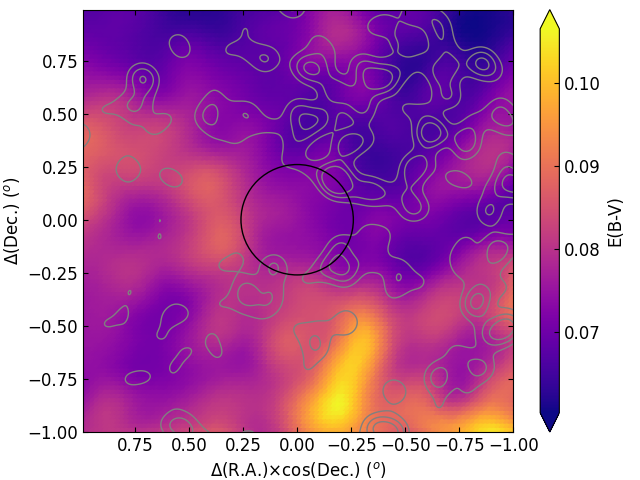}
\caption{Interstellar extinction ($E(B-V)$) map across the observed NGC~6362 field. The
black circle represents the cluster's Jacobi radius (0.26 deg), while the gray contours 
correspond to isodensity levels of the tidal tail stars selected by \citet{kunduetal2019b}.}
\label{fig1}
\end{figure}

\begin{figure}
\includegraphics[width=\columnwidth]{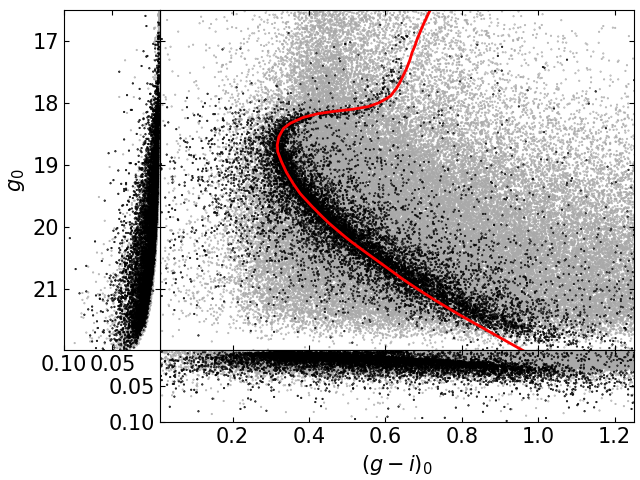}
\caption{CMD for all the measured stars in the field fo NGC~6362, with their respective
photometric errors. Black and gray dots represent stars located inside/outside a circle
of radius equals to 2 times the cluster's half-mass radius (0.11 deg), respectively. 
The red line is an isochrone for the cluster's age (12 Gyr) and metallicity ([Fe/H] = 
-1.07 dex) \citep{gontcharovetal2023}.}
\label{fig2}
\end{figure}

\section{Data analysis}

The region onto which NGC~6362 is projected is affected by interstellar extinction that
varies with the position in the sky. In order to reproduce the interstellar extinction map
towards the cluster, I retrieved the $E(B-V)$ values as a function of R.A. and Dec. 
from \citet{sf11}, provided by the NASA/IPAC Infrared Science Archive\footnote{https://irsa.ipac.caltech.edu/} for the entire analyzed area. Figure~\ref{fig1} illustrates the
resulting extinction map, which reveals that along the cluster's line-of-sight the interstellar 
absorption is relatively low, with a maximum difference between the most and least reddened 
regions of $\Delta$$E(B-V)$ $\la$ 0.05 mag. Based on Figure~\ref{fig1}, I assigned individual 
$E(B-V)$ values to the measured stars according to their positions in the sky. In order to 
correct the observed magnitudes and colors by interstellar extinction, I used the individual 
$E(B-V)$ values and  the $A_\lambda/A_V$ coefficients given by \citet{wch2019}. Figure~\ref{fig2} 
shows the wealth of information that I produced from the observed images, where a nearly 3 mag 
long well-defined cluster's Main Sequence is clearly visible, alongside with the cluster's  Main 
Sequence Turnoff, sub-giant and the fainter segment of the giant branches. In order to highlights 
the cluster's CMD features I used all the measured stars located within a circle of radius equal 
to two times the cluster's half-mass radius \citep[$r_h$=0.11 deg,][]{bv2021}. The positions of 
stars located inside 2$\times$$r_h$ are satisfactorily reproduced by an isochrone \citep{betal12}
for the cluster's age (12 Gyr) and metallicity ([Fe/H] = -1.07 dex) \citep{gontcharovetal2023}.
Figure~\ref{fig2} also shows that the photometric uncertainties of magnitude and color for the 
faintest  observed cluster's stars are smaller than $\sim$ 0.05 mag. NGC~6362 appears projected 
onto a crowded star field (gray dots in Figure~\ref{fig2}).

The strategy chosen to uncover the spatial distribution of stars that formed within NGC~6362
and are now observed beyond its Jacobi radius consists in to statistically identify cluster's
Main Sequence stars located outside that radius, following the recipe outlined by
\citet{zhangetal2022}. For that purpose, I firstly defined a region
along the cluster's Main Sequence, from $g_0$ = 18 mag down to $g_0$ = 21.5 mag, and
secondly, I built a stellar density map for all the stars that fall within that CMD region. 
Unfortunately, this approach is not able to discern on the
cluster membership of these stars. For such an assessment, I need additional information 
such as
proper motions, radial velocities, metallicities, among others. As far as I am aware, 
because of the relative faintness of the involved stars, this piece of information in 
still unavailable. Nevertheless, field stars within the cluster's
Main Sequence region are expected to be distributed throughout the entire field, so that
cluster stars arise as particular shaped stellar overdensities forming an extended envelope, extra-tidal debris 
or tidal tails \citep{pcb2020}.

\begin{figure}
\includegraphics[width=\columnwidth]{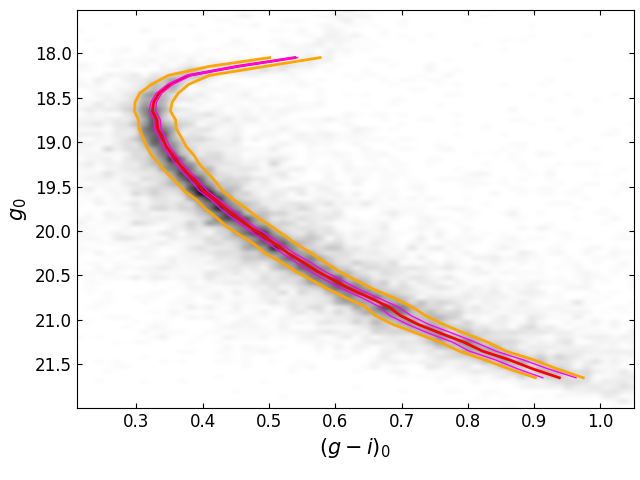}
\caption{Hess diagram for the stars located within 2$\times$$r_h$ from the cluster center.
The traced cluster's Main Sequence ridge line (red), its defined lower and upper color
limits (orange), and those corresponding to the mode of the color error 
distribution (magenta) are superimposed, respectively.}
\label{fig3}
\end{figure}

The cluster's Main Sequence region was devised from the following two steps. I first
traced the cluster's Main Sequence ridge line, and then, I fixed the color width of the
devised region as a function of the magnitude. Both steps were carried out using the
stars located within 2$\times$$r_h$ from the cluster center, i.e., those black points in 
Figure~\ref{fig2}. The cluster's Main Sequence ridge line was determined by computing the 
median of the color distribution for magnitude intervals of $\Delta$$g_0$ = 0.1 mag, while 
their widths correspond to the derived color standard deviations. Figure~\ref{fig3} depicts 
a Hess diagram for the employed stars and the various generated curved, namely: the cluster's
Main Sequence ridge line (red), the lower and upper limits of the cluster's Main Sequence
colors (orange), and the cluster's Main Sequence limits (magenta) corresponding to the mode 
of the color error distribution (see, also, the trend of errors in Figure~\ref{fig2}). As can 
be seen, the photometric errors are notably smaller than the adopted intrinsic width of the
cluster's Main Sequence.

\begin{figure}
\includegraphics[width=\columnwidth]{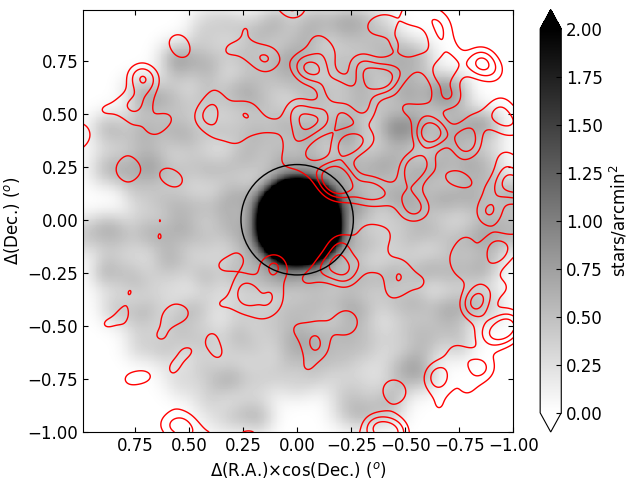}
\caption{Stellar density map of stars distributed within the cluster's Main Sequence
region. The black circle represents the cluster's Jacobi radius, while the red
contours correspond the stellar density levels of stars selected by \citet{kunduetal2019b}
(see Figure~\ref{fig6}).}
\label{fig4}
\end{figure}

\begin{figure}
\includegraphics[width=\columnwidth]{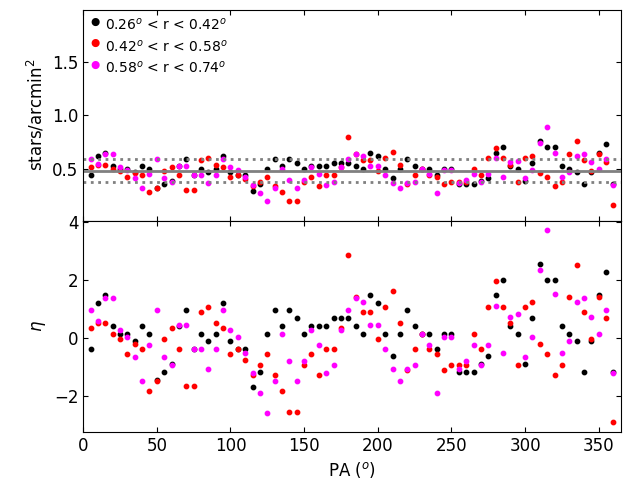}
\caption{Stellar density as a function of the position angle for three different
concentric annulii centered on NGC~6362, as indicated in the top panel. 
The solid and dotted lines represent the mean and standard deviation of all the
plotted points. The significance ($\eta$) of the stellar density over the mean 
background value is depicted in the bottom panel.}
\label{fig5}
\end{figure}

I made use of the \texttt{scikit-learn} software machine learning library \citep{scikit-learn} 
and  its \texttt{Gaussian} kernel density estimator (KDE) to build the stellar density map
for the observed field around NGC~6362, using all the stars distributed within the cluster's
Main Sequence region. I employed a grid of 500$\times$500 
boxes onto the DECam field and used a range of values for the
bandwidth from 0.01$\degr$ up to 0.1$\degr$ in steps of 0.01$\degr$ in order to apply the KDE to 
each generated box. I adopted a bandwidth of 0.05$\degr$ as the optimal 
value, as guided by \texttt{scikit-learn}. Figure~\ref{fig4} shows the resulting density map,
where I used a maximum level of 2 stars arcmin$^{-2}$ in order to highlight lower density levels.
The observed spatial pattern of stars located in the cluster's Main Sequence region shows
a nearly rounded distribution inside the Jacobi radius and no obvious trail of tidal tails
outside it, but a nearly smooth stellar distribution over the entire observed field, as expected 
for a composite  star field population. With the aim of quantifying such a trend, I used three 
different concentric annulii around the cluster center and measured the stellar densities in 
angular sectors of 10 deg wide. The upper panel of Figure~\ref{fig5} illustrates the behavior
of the stellar density as a function of the position angle, measured from North to East.
As can be seen, the three stellar density profiles around the cluster are similar, with
fluctuations that mostly vary around a mean value of 0.48 stars arcmin$^{-2}$ in an amount of
$\sigma$ = 0.11 stars arcmin$^{-2}$. I then computed the deviation from the mean value in the
field in units of the standard deviation, that is,  $\eta$ = (signal $-$ mean value)/standard 
deviation.  From the bottom panel of Figure~\ref{fig5}, I infer that there are some
regions, located particularly in the outer north-western hemisphere, with $\eta$ $\ga$ 2, which
implies a significance level larger than 95$\%$.

\section{Analysis and discussion}

\citet{kunduetal2019b} selected cluster stars ($g_0$ $\la$ 18.0 mag) located beyond its
Jacobi radius based on astrometry filtering criteria. They appear in the  CMD distributed along 
the cluster sub-giant and red giant branches, and horizontal branch as well. Particularly, 
\citet{kunduetal2019b} restricted cluster stars to have proper motions within 3$\sigma$ from the 
mean cluster proper motion. This criterion, which is helpful for identifying cluster stars 
within the cluster body, could fail when applying to stars formed in the cluster that have 
later abandoned it. This is because of tidal tail stars have speed up their pace in order to 
escape the cluster. There are, additionally, other reasons that can contribute to make the
proper motion of tidal tail stars different from the mean cluster
proper motion, among them, projection effects of the tidal tails, 
the Milky Way tidal interaction, the intrinsic kinematic agitation of a 
stellar stream \citep{wanetal2023}. Likewise, it is possible to find field stars projected
along the cluster line of sight that share proper motions similar to cluster stars \citep{piattietal2023}.

\begin{figure}
\includegraphics[width=\columnwidth]{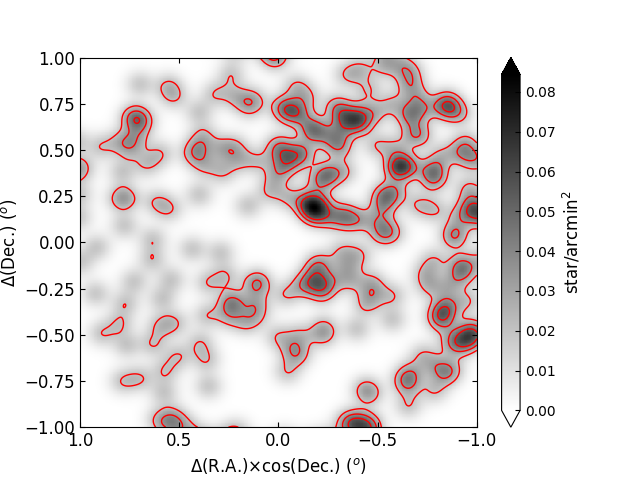}
\caption{Stellar density map for the tidal tail stars selected by \citet{kunduetal2019b},
with contour levels corresponding to 0.02, 0.035, and 0.05 stars arcmin$^2$.}
\label{fig6}
\end{figure}

I used the {\it Gaia} DR2 data \citep{gaiaetal2016,gaiaetal2018b} from  \citet{kunduetal2019b} 
for their selected 259 extra-tidal stars to build a stellar density map following the same
precepts described above to construct Figure~\ref{fig4}. Figure~\ref{fig6} shows the resulting
stellar density map, to which I also traced different contour levels. The stars seem to be 
preferentially distributed across the north-western hemisphere, where I also unveiled
a slightly excess with respect to the south-eastern hemisphere ($\Delta$$\eta$ $\sim$ 2)
for stars located along the cluster's Main Sequence region ($g_0 >$ 18.0 mag). This means that, 
independently of the chosen magnitude range along the cluster CMD, stars distributed 
across the  observed DECam field following a similar spatial pattern. I also note that
Figure~\ref{fig2} reveals a large population of field stars superimposed onto the cluster CMD features (background gray dots). As for comparison purposes, I overplotted the contour 
levels of \citet{kunduetal2019b}'s stars onto
the stellar density map of the cluster's Main Sequence stars (see Figure~\ref{fig4}), and
onto the reddening map (see Figure~\ref{fig1}). Their spatial distribution could suggest that they 
are confined within the lower reddening regions. Nevertheless, the variation of the reddening 
across the field is not so high ($\Delta$$E(B-V)$ $\la$  0.05 mag) as to expect a variation 
in the stellar density of these relatively bright stars due to interstellar absorption.

How to discern whether the spatial distribution of the stars selected by \citet{kunduetal2019b} 
really represent extra-tidal features of NGC~6362? If they were confirmed, then the slightly
stellar excess found from stars distributed within the cluster's Main Sequence region could also 
be an evidence of the presence of such extra-tidal structures. Furthermore, they would be 
very low stellar density tidal tails that do not surpass $\eta$ = 3. According to
\citet{pcb2020}, Milky Milky Way globular clusters can exhibit tidal tails, extended
envelopes or do not show any extra-tidal structures. When I examine the shapes of  
extra-tidal structures of nearly 50 studied globular clusters
\citep[][and references therein]{zhangetal2022}, I conclude that extended envelope
stellar distributions are functions of the distance to the cluster center, no angular
direction arises as preferential. In turn, tidal tails are well collimated stellar structures
that emerge from the cluster body in two directions, namely, the leading and the
trailing tails, respectively. Unfortunately, neither an extended envelope nor tidal tails can be
recognized from the spatial distribution of stars selected by  \citet{kunduetal2019b} 
and in this work (see Figure~\ref{fig4}). 

The presence of tidal tails in some globular clusters
and  the absence of them in others is still a topic of debate. \citet{pcb2020} explored
different kinematical and structural parameter spaces, and found that
globular clusters behave similarly, irrespective of the presence of extended envelopes
or tidal tails, or the absence thereof. \citet{zhangetal2022} showed that globular
clusters with tidal tails or extended envelopes have apogalactocentric distances
$\ga$ 5 kpc, a behavior previously noticed by \citet{piatti2021c}, who suggested that
the lack of detection of tidal tails in bulge globular clusters could be due to the reduced 
diffusion time of tidal tails by the kinematically chaotic nature of the orbits of these 
globular clusters \citep{kunduetal2019b}, thus shortening the time interval during which 
the tidal tails can be detected. Recently, \citet{weatherfordetal2023} re-examined
the behavior of potential escapers in globular clusters dynamically evolving along
chaotic orbits and found diffusion times smaller than 100 Myr. NGC~6362 is
a dynamically evolving globular cluster with a significant internal rotation
\citep{dalessandroetal2021}.

I investigated the possibility that the spatial distribution of \citet{kunduetal2019b}'s 
selected stars (see Figure~\ref{fig6}) can be due to the expected
stellar density of Milky Way field stars along the cluster's line of sight.
For that purpose, I made used of TRILEGAL\footnote{http://stev.oapd.inaf.it/cgi-bin/trilegal} 
\citep{girardietal2005}, a  stellar population synthesis code that  allows
changes in the star formation rate, age-metallicity relation, initial mass function, geometry 
of Milky Way components, among others. I used 81 adjacent squared regions of 
0.0625 deg$^2$ each, that cover uniformly the entire observed DECam field. 
I generated individual CMDs for each sub-field and then counted the number
of stars located within the cluster's Main Sequence region, similarly as we
did to build Figure~\ref{fig4}.

I chose the DECam photometric system, a limiting magnitude of $g,i$ = 23 mag,
and a distance modulus resolution of the Milky Way components of 0.1 mag.
I set the initial mass function according to \citet{kroupa02}, a binary fraction of 0.3 with a
mass ratio larger than 0.7. The interstellar extinction was modeled by an 
exponential disk with a scale height $h_z$ = 100 pc and a scale length $h_R$ = 3200 pc
and a visual absorption variation $\partial$$A_V$/$\partial$$R$ = 0.00015 mag/pc
\citep{lietal2018a}, and the Sun position at $R_{\odot}$ = 8300 pc, and $z_{\odot}$ = 
15 pc \citep{monteiroetal2021}. The Milky Way halo, thin and thick disks, and bulge
were modeled as follows: the halo was represented by a oblate $r^{1/4}$ spheroid with an 
effective radius $r_h$ = 2698.93 pc and an oblateness $q_h$ = 0.583063, and
$\Omega$ = 0.0001 $M_{\odot}$/pc$^3$. The halo star formation rate and
age-metallicity relationship are those given by \citep{rn1991}. The thin disk is
described by an exponential disk along $z$ and $R$, respectively, with a scale height 
increasing with the age: $h_z$ = 94.690$\times$(1 + age/5.55079$\times$10$^9$)$^{1.6666}$,
and a scale length $h_R$ = 2913.36 pc and an outer cutoff at 15000 pc. We
used a 2-step star formation rate, the age-metallicity relationship  with $\alpha$-enrichment
given by \citet{f1998}, and $\Sigma$ = 55.4082 $M_{\odot}$/pc$^2$. The thick disk
is also represented by an exponential disk in both $z$ and $R$ directions, with
scale height $h_z$ =. 800 pc, and scale length $h_R$ = 2394.07 pc and outer
cutoff at 15000 pc; $\Omega$ = 0.001  $M_{\odot}$/pc$^3$. I adopted a
constant star formation rate and $Z$ = 0.008 with $\sigma$[M/H] = 0.1 dex. Finally,
I assumed a triaxial bulge with a scale length of 2500 pc and a truncation
scale length of 95 pc,  and $y/x$ and $z/x$ axial ratios of 0.68 and 0.31, respectively.
The angle between the direction along the bar and that from the Sun to the Milky Way center
was set in 15 deg, and $\Omega$ = 406 $M_{\odot}$/pc$^3$. Its star formation
rate and age-metallicity relationship were taken from \citet{zoccalietal2003}.

\begin{figure}
\includegraphics[width=\columnwidth]{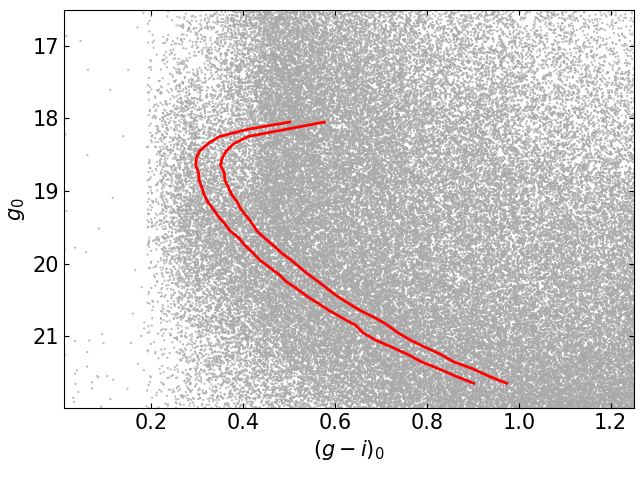}
\caption{Synthetic CMD generated using TRILEGAL for the studied
Milky Way region (see text for details). The red lines represent the devised
cluster's Main Sequence region.}
\label{fig7}
\end{figure}

\begin{figure}
\includegraphics[width=\columnwidth]{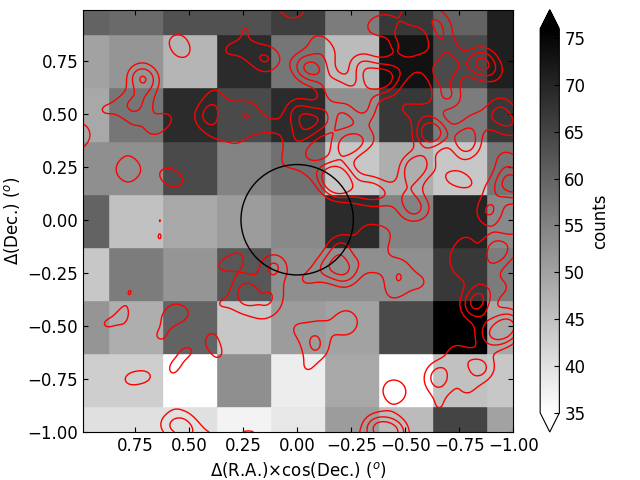}
\caption{Stellar density map built from star counts of stars distributed
within the cluster's Main Sequence region generated  using TRILEGAL 
(see text for details). The red contour levels represent the spatial
distribution of stars selected by \citet{kunduetal2019b}. The black
circle represent the cluster's Jacobi radius.}
\label{fig8}
\end{figure}

Figure~\ref{fig7} illustrates the resulting synthetic CMD for the entire studied field, 
which compares well with that obtained from the observed DECam field
(see Figure~\ref{fig2}); while Figure~\ref{fig8} depicts the stellar density map built from
TRILEGAL stars distributed within the cluster's Main Sequence region. At a 
first glance, there is a similarity between the synthetic stellar density map and the 
spatial distribution of the stars selected by \citet{kunduetal2019b}, which poses
the possibility that they could belong to the composite star field population 
rather than being NGC~6362 tidal tail stars.

\begin{figure}
\includegraphics[width=\columnwidth]{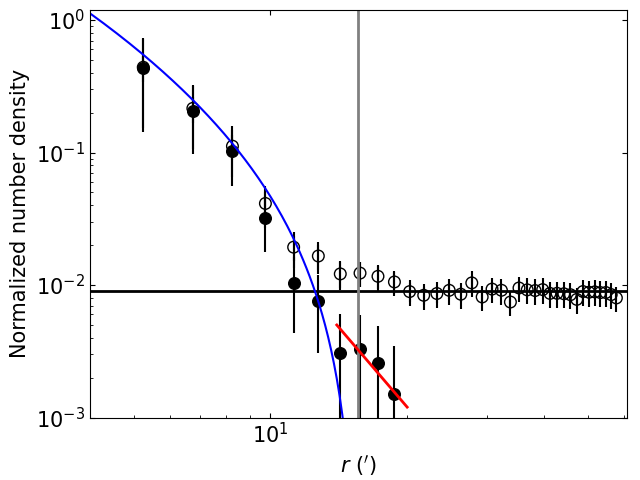}
\caption{Observed and background corrected normalized stellar density profiles
represented by open and filled circles, respectively. Error bars are also shown.
The blue curve represents the King's model for the core and tidal radius of 
the cluster \citep{bh2018}; while the red line corresponds to a power law with
slope equals to 4.0. Black horizontal and gray vertical lines represent the mean
background level and the cluster Jacobi radius, respectively.}
\label{fig9}
\end{figure}

I finally built the cluster stellar density radial profile using the stars distributed within 
the Main Sequence region. I counted the number of stars distributed inside boxes of 0.01 deg up 
to 0.05 deg a side, increasing by steps of 0.01 deg a side, and then built the corresponding radial profiles.
This procedure allowed us to extend the radial profile far from the cluster center, where the area
of the observed DECam field do not cover entirely those of concentric annulii centered on the cluster.
The resulting average radial profile is depicted with open circles in Figure~\ref{fig9}.
As expected, the background level is clearly visibly across an extended region. I then
subtracted to the observed radial profile the mean value of the background stellar
density derived above and superimposed to that background corrected radial profile (filled circles) 
a \citet{king62}'s profile with core and tidal radius taken from \citet{bh2018}. As can be
seen, the \citet{king62}'s profile matches very well the cluster radial profile, which exhibits
around the distance of the cluster's Jacobi radius an excess of stars following a power law distribution
with an slope of 4.0. The latter is remarkably different from the values obtained for Milky Way 
globular clusters with extended envelopes \citep[slope $\sim$ 1; e.g.,][]{olszewskietal2009,p17c}.
Although the
deviation of the radial profile from the \citet{king62}'s profile is visible, I argue that
it does not reveal an extended envelope, but a thin not-uniform low stellar density halo. 
This outcome is in very good agreement with the slightly difference found  between the
north-western and south-eastern outer hemispheres from the stellar 
density map.

\section{Conclusions}

The study of the outermost regions of Milky Way globular clusters is interesting by its own,
since it provides clues about a variety of astrophysical topics for which our understanding 
is far from being complete(see Section ~1). According to a recent study, NGC~6362 presents tidal tails 
out to $\sim$ 2 deg from its center \citep{kunduetal2019b}, which challenges
theoretical predictions for globular clusters in chaotic orbit regimes, and orbiting
close to the Milky Way bulge. I recognized here a science case, and embarked in the first
deep search for extra tidal stars throughout an extended region around the cluster.

The collected data consist of images obtained with the 4.0m Blanco telescope (CTIO) that
reach nearly 3 mag below the cluster's Main Sequence turnoff across an area of $\sim$ 3 deg$^2$, 
a limiting magnitude not achieved by any ground-based observation, but only by HST data
 \citep{dalessandroetal2014} for the innermost cluster region. I scrutinized the  
produced DECam field CMD, from which I selected stars distributed across the entire field and
encompassed within the devised cluster's Main Sequence region. Using those stars we
built a stellar density map, an azimuthal stellar distribution for different concentric
annulii centered on the cluster, and a cluster stellar density radial profile. 
From the stellar density map, I detected an overall relatively smooth stellar density distribution 
from the cluster Jacobi radius out to nearly 3.8$\times$$r_J$, with a slightly difference smaller 
than 2 times the background stellar density fluctuation between the mean stellar density of the 
south-eastern and that of north-western hemispheres, the latter being higher.
The same trend is observed in the plane of the stellar density as a function of the position angle,
and from the stellar radial profile.
I finally constructed a stellar density map from a synthetic CMD generated from a
model of the stellar population distribution in the Milky Way, which also confirms the
above findings.

On the other hand, I produced a stellar density map of the tidal tail stars selected by 
\citet{kunduetal2019b} and superimposed the corresponding contour levels over the present
stellar density map and that obtained from the synthetic CMD. The  \citet{kunduetal2019b}'s
tidal tail stars are distributed mainly throughout the north-western hemisphere, which
suggests that they likely belong to the composite star field population. I also examined
their spatial distribution with that of the interstellar absorption toward the cluster, and
found that they are projected onto the least reddened regions. Nevertheless, I detected
a low stellar density excess around the cluster Jacobi radius, from which
I conclude that NGC~6362 present a thin extra tidal halo. This outcome is two-fold in 
very good agreement with the expected relative high cluster star mass lost by disruption  \citep{piattietal2019b} and the short diffuse time of tidal tails applicable to this 
globular cluster.

\begin{acknowledgements}
I thank the referee for the thorough reading of the manuscript and
timely suggestions to improve it. 

This project used data obtained with the Dark Energy Camera (DECam), which was constructed by 
the Dark Energy Survey (DES) collaboration. Funding for the DES Projects has been provided by the 
US Department of Energy, the US National Science Foundation, the Ministry of Science and Education 
of Spain, the Science and Technology Facilities Council of the United Kingdom, the Higher 
Education Funding Council for England, the National Center for Supercomputing Applications at the
University of Illinois at Urbana-Champaign, the Kavli Institute for Cosmological Physics at the
University of Chicago, Center for Cosmology and Astro-Particle Physics at the Ohio State 
University, the Mitchell Institute for Fundamental Physics and Astronomy at Texas A\&M University, Financiadora de Estudos e Projetos, Funda\c{c}\~{a}o Carlos Chagas Filho de Amparo $\`a$ Pesquisa 
do Estado do Rio de Janeiro, Conselho Nacional de Desenvolvimento Cient\'{\i}fico e Tecnol\'ogico 
and the Minist\'erio da Ci\^{e}ncia, Tecnologia e Inova\c{c}\~{a}o, the Deutsche Forschungsgemeinschaft and the Collaborating Institutions in the Dark Energy Survey.

The Collaborating Institutions are Argonne National Laboratory, the University of California at 
Santa Cruz, the University of Cambridge, Centro de Investigaciones En\'ergeticas, Medioambientales 
y Tecnol\'ogicas–Madrid, the University of Chicago, University College London, the DES-Brazil Consortium, the University of Edinburgh, the Eidgenössische Technische Hochschule (ETH) Zürich, 
Fermi National Accelerator Laboratory, the University of Illinois at Urbana-Champaign, the 
Institut de Ci\`encies de l’Espai (IEEC/CSIC), the Institut de F\'{\i}sica d’Altes Energies, 
Lawrence Berkeley National Laboratory, the Ludwig-Maximilians Universit\"at München and the
associated Excellence Cluster Universe, the University of Michigan, NSF’s NOIRLab, the University 
of Nottingham, the Ohio State University, the OzDES Membership Consortium, the University of Pennsylvania, the University of Portsmouth, SLAC National Accelerator Laboratory, Stanford University, the University of Sussex, and Texas A\&M University.

Based on observations at Cerro Tololo Inter-American Observatory, NSF’s NOIRLab (NOIRLab Prop. ID 
2023A-627924; PI: A. Piatti), which is managed by the Association of Universities for Research in Astronomy (AURA) under a cooperative agreement with the National Science Foundation.

Data for reproducing the figures and analysis in this work will be available upon request
to the author.

\end{acknowledgements}


\begin{thebibliography}{41}
\expandafter\ifx\csname natexlab\endcsname\relax\def\natexlab#1{#1}\fi

\bibitem[{{Baumgardt} \& {Hilker}(2018)}]{bh2018}
{Baumgardt}, H. \& {Hilker}, M. 2018, \mnras, 478, 1520

\bibitem[{{Baumgardt} \& {Vasiliev}(2021)}]{bv2021}
{Baumgardt}, H. \& {Vasiliev}, E. 2021, \mnras, 505, 5957

\bibitem[{{Bressan} {et~al.}(2012){Bressan}, {Marigo}, {Girardi}, {Salasnich},
  {Dal Cero}, {Rubele}, \& {Nanni}}]{betal12}
{Bressan}, A., {Marigo}, P., {Girardi}, L., {et~al.} 2012, \mnras, 427, 127

\bibitem[{{Carballo-Bello} {et~al.}(2012){Carballo-Bello}, {Gieles}, {Sollima},
  {Koposov}, {Mart{\'{\i}}nez-Delgado}, \&
  {Pe{\~n}arrubia}}]{carballobelloetal2012}
{Carballo-Bello}, J.~A., {Gieles}, M., {Sollima}, A., {et~al.} 2012, \mnras,
  419, 14

\bibitem[{{Carballo-Bello} {et~al.}(2014){Carballo-Bello}, {Sollima},
  {Mart{\'\i}nez-Delgado}, {Pila-D{\'\i}ez}, {Leaman}, {Fliri}, {Mu{\~n}oz}, \&
  {Corral-Santana}}]{carballobelloetal2014}
{Carballo-Bello}, J.~A., {Sollima}, A., {Mart{\'\i}nez-Delgado}, D., {et~al.}
  2014, \mnras, 445, 2971

\bibitem[{{Carlberg} \& {Grillmair}(2021)}]{cg2021}
{Carlberg}, R.~G. \& {Grillmair}, C.~J. 2021, \apj, 922, 104

\bibitem[{{Dalessandro} {et~al.}(2014){Dalessandro}, {Massari}, {Bellazzini},
  {Miocchi}, {Mucciarelli}, {Salaris}, {Cassisi}, {Ferraro}, \&
  {Lanzoni}}]{dalessandroetal2014}
{Dalessandro}, E., {Massari}, D., {Bellazzini}, M., {et~al.} 2014, \apjl, 791,
  L4

\bibitem[{{Dalessandro} {et~al.}(2021){Dalessandro}, {Raso}, {Kamann},
  {Bellazzini}, {Vesperini}, {Bellini}, \& {Beccari}}]{dalessandroetal2021}
{Dalessandro}, E., {Raso}, S., {Kamann}, S., {et~al.} 2021, \mnras, 506, 813

\bibitem[{{de Boer} {et~al.}(2019){de Boer}, {Gieles}, {Balbinot},
  {H{\'e}nault-Brunet}, {Sollima}, {Watkins}, \& {Claydon}}]{deboeretal2019}
{de Boer}, T.~J.~L., {Gieles}, M., {Balbinot}, E., {et~al.} 2019, \mnras, 485,
  4906

\bibitem[{{Flaugher} {et~al.}(2015){Flaugher}, {Diehl}, {Honscheid}, {Abbott},
  {Alvarez}, {Angstadt}, {Annis}, {Antonik}, {Ballester}, {Beaufore},
  {Bernstein}, {Bernstein}, {Bigelow}, {Bonati}, {Boprie}, {Brooks},
  {Buckley-Geer}, {Campa}, {Cardiel-Sas}, {Castander}, {Castilla}, {Cease},
  {Cela-Ruiz}, {Chappa}, {Chi}, {Cooper}, {da Costa}, {Dede}, {Derylo},
  {DePoy}, {de Vicente}, {Doel}, {Drlica-Wagner}, {Eiting}, {Elliott}, {Emes},
  {Estrada}, {Fausti Neto}, {Finley}, {Flores}, {Frieman}, {Gerdes},
  {Gladders}, {Gregory}, {Gutierrez}, {Hao}, {Holland}, {Holm}, {Huffman},
  {Jackson}, {James}, {Jonas}, {Karcher}, {Karliner}, {Kent}, {Kessler},
  {Kozlovsky}, {Kron}, {Kubik}, {Kuehn}, {Kuhlmann}, {Kuk}, {Lahav}, {Lathrop},
  {Lee}, {Levi}, {Lewis}, {Li}, {Mandrichenko}, {Marshall}, {Martinez},
  {Merritt}, {Miquel}, {Mu{\~n}oz}, {Neilsen}, {Nichol}, {Nord}, {Ogando},
  {Olsen}, {Palaio}, {Patton}, {Peoples}, {Plazas}, {Rauch}, {Reil}, {Rheault},
  {Roe}, {Rogers}, {Roodman}, {Sanchez}, {Scarpine}, {Schindler}, {Schmidt},
  {Schmitt}, {Schubnell}, {Schultz}, {Schurter}, {Scott}, {Serrano}, {Shaw},
  {Smith}, {Soares-Santos}, {Stefanik}, {Stuermer}, {Suchyta}, {Sypniewski},
  {Tarle}, {Thaler}, {Tighe}, {Tran}, {Tucker}, {Walker}, {Wang}, {Watson},
  {Weaverdyck}, {Wester}, {Woods}, {Yanny}, \& {DES
  Collaboration}}]{flaugheretal2015}
{Flaugher}, B., {Diehl}, H.~T., {Honscheid}, K., {et~al.} 2015, \aj, 150, 150

\bibitem[{{Fuhrmann}(1998)}]{f1998}
{Fuhrmann}, K. 1998, \aap, 338, 161

\bibitem[{{Gaia Collaboration} {et~al.}(2018){Gaia Collaboration}, {Brown},
  {Vallenari}, {Prusti}, {de Bruijne}, {Babusiaux}, {Bailer-Jones}, {Biermann},
  {Evans}, {Eyer}, \& et~al.}]{gaiaetal2018b}
{Gaia Collaboration}, {Brown}, A.~G.~A., {Vallenari}, A., {et~al.} 2018, \aap,
  616, A1

\bibitem[{{Gaia Collaboration} {et~al.}(2016){Gaia Collaboration}, {Prusti},
  {de Bruijne}, {Brown}, {Vallenari}, {Babusiaux}, {Bailer-Jones}, {Bastian},
  {Biermann}, {Evans}, \& et~al.}]{gaiaetal2016}
{Gaia Collaboration}, {Prusti}, T., {de Bruijne}, J.~H.~J., {et~al.} 2016,
  \aap, 595, A1

\bibitem[{{Girardi} {et~al.}(2005){Girardi}, {Groenewegen}, {Hatziminaoglou},
  \& {da Costa}}]{girardietal2005}
{Girardi}, L., {Groenewegen}, M.~A.~T., {Hatziminaoglou}, E., \& {da Costa}, L.
  2005, \aap, 436, 895

\bibitem[{{Gnedin} \& {Ostriker}(1997)}]{go1997}
{Gnedin}, O.~Y. \& {Ostriker}, J.~P. 1997, \apj, 474, 223

\bibitem[{{Gontcharov} {et~al.}(2023){Gontcharov}, {Khovritchev}, {Mosenkov},
  {Il'in}, {Marchuk}, {Poliakov}, {Ryutina}, {Savchenko}, {Smirnov}, {Usachev},
  {Lee}, {Camacho}, \& {Hebdon}}]{gontcharovetal2023}
{Gontcharov}, G.~A., {Khovritchev}, M.~Y., {Mosenkov}, A.~V., {et~al.} 2023,
  \mnras, 518, 3036

\bibitem[{{Hozumi} \& {Burkert}(2015)}]{hb2015}
{Hozumi}, S. \& {Burkert}, A. 2015, \mnras, 446, 3100

\bibitem[{{King}(1962)}]{king62}
{King}, I. 1962, \aj, 67, 471

\bibitem[{Kroupa(2002)}]{kroupa02}
Kroupa, P. 2002, Science, 295, 82

\bibitem[{{Kundu} {et~al.}(2019){Kundu}, {Fern{\'a}ndez-Trincado}, {Minniti},
  {Singh}, {Moreno}, {Reyl{\'e}}, {Robin}, \& {Soto}}]{kunduetal2019b}
{Kundu}, R., {Fern{\'a}ndez-Trincado}, J.~G., {Minniti}, D., {et~al.} 2019,
  \mnras, 489, 4565

\bibitem[{{Li} {et~al.}(2018){Li}, {Shen}, {Hou}, {Yuan}, {Xiang}, {Chen},
  {Huang}, \& {Liu}}]{lietal2018a}
{Li}, L., {Shen}, S., {Hou}, J., {et~al.} 2018, \apj, 858, 75

\bibitem[{{Mackey} {et~al.}(2019){Mackey}, {Ferguson}, {Huxor}, {Veljanoski},
  {Lewis}, {McConnachie}, {Martin}, {Ibata}, {Irwin}, {C{\^o}t{\'e}},
  {Collins}, {Tanvir}, \& {Bate}}]{mackeyetal2019}
{Mackey}, A.~D., {Ferguson}, A.~M.~N., {Huxor}, A.~P., {et~al.} 2019, \mnras,
  484, 1756

\bibitem[{{Mestre} {et~al.}(2020){Mestre}, {Llinares}, \&
  {Carpintero}}]{mestreetal2020}
{Mestre}, M., {Llinares}, C., \& {Carpintero}, D.~D. 2020, \mnras, 492, 4398

\bibitem[{{Monteiro} {et~al.}(2021){Monteiro}, {Barros}, {Dias}, \&
  {L{\'e}pine}}]{monteiroetal2021}
{Monteiro}, H., {Barros}, D.~A., {Dias}, W.~S., \& {L{\'e}pine}, J. R.~D. 2021,
  Frontiers in Astronomy and Space Sciences, 8, 62

\bibitem[{{Olszewski} {et~al.}(2009){Olszewski}, {Saha}, {Knezek},
  {Subramaniam}, {de Boer}, \& {Seitzer}}]{olszewskietal2009}
{Olszewski}, E.~W., {Saha}, A., {Knezek}, P., {et~al.} 2009, \aj, 138, 1570

\bibitem[{Pedregosa {et~al.}(2011)Pedregosa, Varoquaux, Gramfort, Michel,
  Thirion, Grisel, Blondel, Prettenhofer, Weiss, Dubourg, Vanderplas, Passos,
  Cournapeau, Brucher, Perrot, \& Duchesnay}]{scikit-learn}
Pedregosa, F., Varoquaux, G., Gramfort, A., {et~al.} 2011, Journal of Machine
  Learning Research, 12, 2825

\bibitem[{{Piatti}(2017)}]{p17c}
{Piatti}, A.~E. 2017, \mnras, 466, 4960

\bibitem[{{Piatti}(2021)}]{piatti2021c}
{Piatti}, A.~E. 2021, \mnras, 505, 3033

\bibitem[{{Piatti} \& {Carballo-Bello}(2020)}]{pcb2020}
{Piatti}, A.~E. \& {Carballo-Bello}, J.~A. 2020, \aap, 637, L2

\bibitem[{{Piatti} {et~al.}(2023){Piatti}, {Illesca}, {Massara}, {Chiarpotti},
  {Rold{\'a}n}, {Mor{\'o}n}, \& {Bazzoni}}]{piattietal2023}
{Piatti}, A.~E., {Illesca}, D. M.~F., {Massara}, A.~A., {et~al.} 2023, \mnras,
  518, 6216

\bibitem[{{Piatti} {et~al.}(2019){Piatti}, {Webb}, \&
  {Carlberg}}]{piattietal2019b}
{Piatti}, A.~E., {Webb}, J.~J., \& {Carlberg}, R.~G. 2019, \mnras, 489, 4367

\bibitem[{{Ryan} \& {Norris}(1991)}]{rn1991}
{Ryan}, S.~G. \& {Norris}, J.~E. 1991, \aj, 101, 1865

\bibitem[{{Schlafly} \& {Finkbeiner}(2011)}]{sf11}
{Schlafly}, E.~F. \& {Finkbeiner}, D.~P. 2011, \apj, 737, 103

\bibitem[{{Starkman} {et~al.}(2020){Starkman}, {Bovy}, \&
  {Webb}}]{starkmanetal2020}
{Starkman}, N., {Bovy}, J., \& {Webb}, J.~J. 2020, \mnras, 493, 4978

\bibitem[{{Stetson} {et~al.}(1990){Stetson}, {Davis}, \& {Crabtree}}]{setal90}
{Stetson}, P.~B., {Davis}, L.~E., \& {Crabtree}, D.~R. 1990, in Astronomical
  Society of the Pacific Conference Series, Vol.~8, CCDs in astronomy, ed.
  G.~H. {Jacoby}, 289--304

\bibitem[{{Wan} {et~al.}(2023){Wan}, {Arnold}, {Oliver}, {Lewis}, {Baumgardt},
  {Gieles}, {H{\'e}nault-Brunet}, {de Boer}, {Balbinot}, {Da Costa}, {Mackey},
  {Erkal}, {Ferguson}, {Kuzma}, {Pancino}, {Pe{\~n}arrubia}, {Sanna},
  {Sollima}, {van der Marel}, \& {Watkins}}]{wanetal2023}
{Wan}, Z., {Arnold}, A.~D., {Oliver}, W.~H., {et~al.} 2023, \mnras, 519, 192

\bibitem[{{Wan} {et~al.}(2021){Wan}, {Oliver}, {Baumgardt}, {Lewis}, {Gieles},
  {H{\'e}nault-Brunet}, {de Boer}, {Balbinot}, {Da Costa}, \&
  {Mackey}}]{wanetal2021}
{Wan}, Z., {Oliver}, W.~H., {Baumgardt}, H., {et~al.} 2021, \mnras, 502, 4513

\bibitem[{{Wang} \& {Chen}(2019)}]{wch2019}
{Wang}, S. \& {Chen}, X. 2019, \apj, 877, 116

\bibitem[{{Weatherford} {et~al.}(2023){Weatherford}, {Rasio}, {Chatterjee},
  {Fragione}, {K{\i}ro{\u{g}}lu}, \& {Kremer}}]{weatherfordetal2023}
{Weatherford}, N.~C., {Rasio}, F.~A., {Chatterjee}, S., {et~al.} 2023, arXiv
  e-prints, arXiv:2310.01485

\bibitem[{{Zhang} {et~al.}(2022){Zhang}, {Mackey}, \& {Da
  Costa}}]{zhangetal2022}
{Zhang}, S., {Mackey}, D., \& {Da Costa}, G.~S. 2022, \mnras, 513, 3136

\bibitem[{{Zoccali} {et~al.}(2003){Zoccali}, {Renzini}, {Ortolani}, {Greggio},
  {Saviane}, {Cassisi}, {Rejkuba}, {Barbuy}, {Rich}, \&
  {Bica}}]{zoccalietal2003}
{Zoccali}, M., {Renzini}, A., {Ortolani}, S., {et~al.} 2003, \aap, 399, 931

\end{thebibliography}


\end{document}